Probing thermal waves on the free surface of various media: Surface Fluctuation Specular Reflection Spectroscopy

A. Tay[a], C. Thibierge,[a] D. Fournier,[b] C. Fretigny,[a] F. Lequeux,[a] C. Monteux,[a] J.P. Roger[b] and L. Talini [a].

[a] Laboratoire PPMD, UMR 7615 CNRS/UPMC and ESPCI, 10 rue Vauquelin 75005 Paris, France.

[b] Laboratoire PEM, UPR A0005 CNRS/UPMC and ESPCI , 10 rue Vauquelin, 75005 Paris, France.

Abstract : Thermal motion gives rise to fluctuations in free surfaces; the propagation of the thermally excited waves on such surfaces depends on the mechanical properties of the medium. Their measurement can therefore provide information on those properties. We have developed an optical tool to probe the thermally excited waves on free surfaces: Surface Fluctuation Specular Reflection (SFSR) spectroscopy. It consists in measuring the fluctuations in the position of a laser beam, which is specularly reflected onto the free surface of a medium, and is therefore sensitive to the roughness of that surface. We show how the measured signal is related to the medium properties. We also present measurements performed on Newtonian liquids as well as on a viscoelastic solid; we show that, in all cases, there is a very good agreement between experimental and computed spectra. SFSR thus applies to a broad range of materials. It moreover offers a very good temporal resolution and should provide a useful tool for dynamical measurements on complex fluids.

**I INTRODUCTION**

Measuring the effect of thermal fluctuations can provide a measurement of the mechanical properties of a material. In the case of complex fluids, characterization based upon thermal fluctuations measurements presents several advantages as compared with, for instance, conventional macroscopic rheology; in particular, the measurements are local and no externally induced flow is needed. Various techniques using the thermal fluctuations of tracers either embedded in a medium[1-2] or in contact with its surface[3] have thus been developed. They provide successful measurements of the viscoelastic properties of polymer gels and biological materials over a broad frequency range. However, those techniques require the addition of tracers that can for instance induce modifications in the structure of the medium; moreover, the probed scale is limited by the tracer size that must be kept close to 1µm.

Other techniques, that present the advantage to be non-invasive, have been developed in the past. Their principle is to measure the thermally excited waves at free surfaces. Those waves being governed by surface energy, viscoelastic or inertial effects according to the probed frequency, such measurements can therefore be related to the properties of the material. For usual values of interface energy, the amplitude of thermal waves is very small, the roughness of the interface being of the order of a few tenths of a nanometer. It is nevertheless possible to detect them using such techniques as surface scattering, ellipsometry or reflectivity[4]. In the first technique, light or X-ray scattering provides dynamical information on the thermal fluctuations[1] and constitutes a very powerful way to measure the surface mechanical properties – surface viscoelasticity – for instance of polymer films[5-6] or surfactant monolayers.[4,7-8] In the case of media with a bulk viscoelasticity, previous experimental attempts have not proved successful[9] because of the low intensity of the measured signals. On the other hand, ellipsometry as well as reflectivity techniques[10-11] only provide a measurement of the time averaged roughness of a free surface, and neither the time nor spatial variations of its thermal fluctuations.

We present herein a technique that allows detection of the thermally excited waves on free surfaces: Surface Fluctuation Specular Reflection (SFSR) spectroscopy. SFSR consists in measuring the position of a laser beam that is specularly reflected from the considered free surface. We show that the thermally excited waves on the free surfaces of various Newtonian liquids can thus be detected in a broad frequency range, and that the technique also successfully applies to a medium exhibiting bulk viscoelasticity, a rubber sample. SFSR is therefore well suited for dynamical measurements on complex fluids exhibiting a broad range of mechanical properties.

In section II we detail the experimental set-up. The detected signal and how it is related to the properties of the probed medium are specified in section III and, in sections IV and V, we report the experimental measurements performed with respectively simple liquids and a viscoelastic solid and, in each case, they are compared with the expected variations computed in section III.

## II EXPERIMENTAL SET-UP

SFSR consists in measuring the deflection of a laser beam that is specularly reflected from the sample free surface. To a first approximation, the beam position is sensitive to the slope of that interface (Fig. 1). The experimental set-up is schematized in Fig. 2.

The He-Ne laser beam is focused on the interface using a converging lens; a mirror directs light onto the interface with an almost normal incidence. The radius $R$ of the beam at the beam waist (that coincides with the air-liquid interface) has been set to 30 µm for the experiments we present herein. A decrease of $R$ leads to an increase in the number of spatial modes contributing to the signal; as shown in the following, the nature of the effects probed in the available frequency range also depends on $R$, its value must therefore be chosen according to the system probed.

If liquid, the sample is placed in a plexiglass cell (diameter 5cm) that has been specially designed to eliminate parasitic light (induced by further reflections of the beam) away from the reflected beam. The liquid depth is kept small enough (close to 1mm) to minimize low frequency noise, through an efficient damping of the mechanically induced vibrations. The liquid depth however remains large as compared to the largest wavelengths probed, since, as shown in the following, the contributions to the signal of the large-wavelength modes is small.

The instantaneous deviations of the reflected beam are measured using the two photodiodes of a position sensor. Prior to each measurement, the position sensor is centered with respect to the reflected beam, such that the average light intensities on both photodiodes are identical, i.e. $I_+ = I_-$, where $I_+$ and $I_-$ are respectively the light intensities collected by the upper and lower photodiodes of the position sensor. Owing to the different reflexions, the total light intensity $I_+ + I_-$ is less than 20% of the laser intensity. We denote as δ$I$ the difference of the light intensities, $\delta I = I_+ - I_-$. The time fluctuations of δ$I$ are measured through the changes in the difference of photodiode voltages δ$V$, obtained with a current to voltage converter. Voltage difference δ$V$ is further anti-aliasing filtered, amplified and digitized. The duration of one run as well as the sampling rate can be varied; typical values for the experiments at high frequencies are 2.6s for the duration of a run and $2 \times 10^5$ Hz for the sampling rate. The Fourier transform of δ$V$ is then computed and its squared modulus is averaged over a given number of runs. We denote as $S(\omega)$ the resulting signal that has been normalized by the average voltage delivered by both photodiodes. We show in section III how $S(\omega)$ is related to the surface waves and how it provides a measurement of the properties of the probed medium.

Special care has been taken to reduce noise whether mechanical or electrical in the investigated frequency range (from 0.1 Hz to 100 kHz). In particular, the 4mW He-Ne laser (wavelength 632.8nm) has been selected for its stability. Air convection currents constitute a major source of noise. Those effects are reduced by enclosing both the laser path and the whole set-up. Otherwise, mechanical noise control is not as critical as in scattering experiments,[4] the signal being much more intense than the one in the latter experiments. Finally, the typical duration of one measurement is of the order of 10 min, liquid evaporation during the experiment can therefore be neglected.

At a wavelength comparable to the beam size, wave amplitudes of a few picometers correspond to an amplified voltage difference delivered by the photodiodes of 10 mV, they are therefore easily detectable. The roughness of a broad range of interfaces can therefore be measured, its value being rather of a few hundreds of picometers for instance for usual air/liquid interfaces. Practically, the accuracy strongly depends on the medium probed. At small frequencies, it can be limited by mechanical noise for low viscosity liquids. As shown in section IV, the smallest measurement frequency depends on the liquid viscosity (the mechanical vibrations are less damped by low viscosity fluids), but also on the surface tension (the signal being larger for small surface tensions). At higher frequencies, the noise of the acquisition system and of the position sensor limits the measurement and only signals larger than $2 \times 10^{-13}$ rad$^2$/Hz can be accurately measured.

The experimental set up we present is very similar to the one that has been developed to measure the propagation of excited waves at free surfaces of liquids: Waves are detected using the deflection of a laser beam focused at the surface.[12-14] However, in the latter experiment, only the amplitude of the wave at a given frequency is measured. The originality of SFSR is, first, to deal with thermally excited waves and, second, to detect the contributions from different surface modes. Such a detection allows measurements in a broad frequency range (over up to five decades in frequency) and on systems for which the amplitude of thermal waves is small.

In the following section we show how the signal we detect is related to the properties of the medium probed.

## III DETECTED SIGNAL

The measured signal $S(\omega)$ corresponds to the power spectrum density of the normalised intensity fluctuations i.e., following the Wiener-Khintchine theorem, to the Fourier transform of the time correlation function of $\delta I$ :

$$S(\omega) = \frac{1}{I_0^2} \int_{-\infty}^{\infty} dt \langle \delta I(\tau) \delta I^*(t+\tau) \rangle_\tau e^{-i\omega t} \quad (1)$$

$I_0$ being the total intensity on the photodiode. The angular brackets in equation (1) denote the average over time $\tau$.

We show in the appendix that the measured signal $S(\omega)$ can be related to the power spectrum of the height of the interface, $P_k(\omega)$, such that:

$$S(\omega) = \int_0^\infty P_k(\omega) \Phi(k) k \, dk \quad (2)$$

The summation over all the wavevector moduli $k$ results from the contribution to the signal of several spatial modes of the free surface. This is indeed different from scattering experiments in which a narrow range of spatial modes is selected. The function $\Phi(k)$ in equation (2) reflects the unequal contribution to the signal of all the spatial modes. Whereas the shortest wavelengths contribute to light scattering, the beam deflection is mainly sensitive to surface waves with wavelength of the same order as or larger than the beam size. Therefore, $\Phi(k)$, whose expression is given in the appendix, is a function of the laser beam characteristics, i.e. size at beam waist and wavelength $\lambda$. It can be well approximated by equation (A12).

The power spectrum of the height of the interface, $P_k(\omega)$, introduced in equation (2), is defined as:

$$P_k(\omega) = \frac{A}{2\pi} \int dt e^{-i\omega t} \left\langle h(\vec{k};t) h^*(\vec{k};t+\tau) \right\rangle_\tau \quad (3)$$

Here $h(\vec{k};t)$ is the two dimensional spatial Fourier transform of $h(\vec{r};t)$, the height of the free interface of the liquid at position $\vec{r}$ in the plane of the gas-liquid interface and time $t$: $h(\vec{k};t) = \frac{1}{A} \iint d\vec{r} h(\vec{r};t) e^{i\vec{k}\cdot\vec{r}}$ and $A$ is the surface area of the sample. $h^*(\vec{k};t)$ is the complex conjugate of $h(\vec{k};t)$.

The spectrum $P_k(\omega)$ can be computed by solving the linearized Navier-Stokes equations.[15] In the case of the free surface of a liquid of viscosity $\eta$, density $\rho$ and liquid-gas interfacial energy $\sigma$, $P_k(\omega)$ is given by:

$$P_k(\omega) = \frac{k_b T}{\pi \omega} Im\left(\frac{1}{z}\right) \quad (4)$$

where $k_B T$ is the thermal energy and $z$ is such that:

$$z = \sigma k^2 - i\omega\left(\frac{\eta m(k+m)}{k}\right) + i\omega\eta \frac{(k-m)^2}{k+m} \quad (5)$$

with:

$$m = \left[k^2 - \frac{i\omega\rho}{\eta}\right]^{1/2} \quad (6)$$

These expressions have been derived from more general expressions[4] in the particular case of negligible effects from the gas phase and an infinite depth of the sample. Equation (4) is an expression for the fluctuation-dissipation theorem and is thus valid for systems in thermodynamic equilibrium.

In summary, the measured signal $S(\omega)$, whose general expression is given by equation (2), is related to the laser beam parameters through the function $\Phi(k)$ and to the properties of the sample through the power spectrum $P_k(\omega)$ (equation (4)). In the next two sections, we give that relation in the case of a simple liquid of constant viscosity $\eta$ and in the case of a viscoelastic medium of complex viscosity $\eta(\omega)$. We will show that, in both cases, the given expressions are in very good agreement with the experimental measurements.

**IV SIMPLE LIQUIDS**

The spectrum of the capillary waves on the surface of a Newtonian liquid depends on the relative importance of surface tension, viscous and inertial effects. Using the expressions given in section III, the signal we measure in the case of a Newtonian liquid of viscosity $\eta$, density $\rho$ and surface tension $\sigma$, can be written as:

$$S(\omega) = 2b \frac{k_B T}{\lambda^2} \frac{\rho R}{\eta^2} \int_0^\infty \frac{x^4 \left( \frac{\omega_1}{\omega} x \, Im(\tilde{m}) + 1 \right) e^{-\frac{x^2}{c}}}{\left[ x^3 + x^2 \left( \frac{\omega_2}{4\omega_1} - Re(\tilde{m}) \right) - \frac{1}{4x} \left( \frac{\omega}{\omega_1} \right)^2 \right]^2 + x^2 \left( x \, Im(\tilde{m}) + \frac{\omega}{\omega_1} \right)^2} dx \quad (7)$$

where $Im(\tilde{m})$ and $Re(\tilde{m})$ denote respectively the imaginary and real parts of $\tilde{m} = mR = \left[ x^2 - i\,\omega/\omega_1 \right]^{1/2}$ with $x = kR$. The numerical constants $b$ and $c$ are the ones involved in the expression for function $\Phi$ given by equation (A12).

Frequencies $\omega_1$ and $\omega_2$, that appear in equation (7), depend on the liquid properties and beam size; they are given by $\omega_1 = \eta/\rho R^2$ and $\omega_2 = \sigma/\eta R$. The shape of the spectrum $S(\omega)$ only depends on the values of the frequency $\omega_1$ and on the ratio $\omega_2/\omega_1$, which is also the ratio of the Reynolds and capillary numbers at scale $R$. In the investigated frequency range, the nature of the effects governing the propagation of thermal waves therefore depends in particular on the beam radius.

In the case of relatively viscous fluids exhibiting typical values of surface tension, the frequencies $\omega_1$ and $\omega_2$ are such that $\omega_1 \gg \omega_2$. From equation (7), one can show that at frequencies smaller than $\omega_2$, $S(\omega)$ is roughly constant and scales as $\eta R/\sigma^2$. Whereas at intermediate frequencies ($\omega_1 \gg \omega \gg \omega_2$), it varies inversely with the square of the frequency: $S(\omega) \propto 1/\eta R \omega^2$. Above the frequency $\omega_1$, the decrease of $S(\omega)$ with the frequency also follows a power law but with an exponent -4: $S(\omega) \propto \eta/\rho^2 R^5 \omega^4$.

Those behaviours are observed on the spectra of Fig. 3 that have been obtained experimentally with Newtonian silicon oils of calibrated viscosities (Rhodorsil, 47V20, 47V100, 47V350 and 47V1000). The three regimes we describe are observed in the case of the less viscous oil ($\eta$=20 mPa.s), for which $\omega_1/2\pi$ is of a few kHz, i.e. is encompassed in the spanned frequency range. In the cases of the more viscous oils ($\eta$=100 mPa.s, 350 mPa.s and 1000 mPa.s), only the first two regimes are visible. In any case, the experimental data is compared to the theoretical values of $S(\omega)$ given by equation (7) that has been integrated numerically. The agreement between experimental and theoretical variations is excellent. The computed values of equation (7) have been multiplied by a numerical factor of 7±1 for all experiments in order to adjust the theoretical amplitudes with the experimental ones. The curve shapes are nevertheless described without any adjustable parameter, the viscosities and surface tensions of the oils being given.

When, on the contrary, frequency $\omega_1$ is smaller than $\omega_2$, the shape of $S(\omega)$ changes: at small frequencies, $S(\omega)$ is an increasing function whereas at large frequencies it follows a power law with an exponent that can be smaller than -4. At even larger frequencies however, inertial effects dominate and the decrease in $\omega^{-4}$ is found again. The two former behaviours are illustrated in Fig. 4 in which we have reported the spectrum measured at an interface air/water. In that case, $\omega_1/2\pi$=200Hz and $\omega_2/2\pi$=170kHz. The experimental data is fully described by the expression we have computed for $S(\omega)$, with a surface tension of 60mN/m. The relatively small surface tension measured for the air/water interface may result from the presence of contaminants since no special care was taken to keep the surface clean. As in the case of the less viscous silicon oil, the measurement is limited at small frequencies by mechanical noise. These results nevertheless show that, even in the unfavourable case of a low viscosity and large surface tension liquid such as water, the measurements can be performed over more than two decades in frequency.

Finally, the spatial modes $k$ probed with those liquids can be determined by studying the variations of the integrand in equation (7) as a function of $x=kR$. Whatever the nature of the liquid, the smallest wavelength probed is given by the beam radius and ($\lambda_{min} \approx 2R$), whereas the largest wavelength that significantly contributes to the signal depends on the sample. The modes with a significant contribution nevertheless correspond to wavelengths smaller than the liquid depth, which is close to 1 mm.

In conclusion, we have made successful measurements on interfaces air/liquids whose surface tensions range from about 20mN/m to 60mN/m and whose viscosities range from 1mPa.s to 1000mPa.s. The accuracy of the measurement depends on the fluid properties at small frequencies whereas it is limited by the set-up at higher frequencies. We show that in the large range of fluid properties investigated herein, the measurements can be made over two decades in frequency in the most unfavourable case, and up to five decades in the other cases.

## V VISCOELASTIC MEDIUM

In the case of a viscoelastic medium, the power spectrum can be derived by simply substituting the viscosity η in equation (5) with the complex viscosity of the medium:[16]

$$\eta(\omega) = \frac{1}{\omega}(G''(\omega) - iG'(\omega)) \quad (8)$$

where $G'(\omega)$ and $G''(\omega)$ are respectively the frequency dependent elastic and loss moduli of the medium.

Within the investigated frequency range, the regimes we probe are driven by viscoelasticity and not by capillary effects. The elastic mode regime corresponds to $\omega \ll \omega'_1 = \sqrt{|G|/\rho}/R$ where $|G|$ is the minimum value of $G(\omega) = \sqrt{G'^2(\omega) + G''^2(\omega)}$ ; in that frequency range, the signal $S(\omega)$ can be approximated such that:

$$S(\omega) = d \frac{k_B T}{\lambda^2} \frac{1}{R\omega} \frac{G''(\omega)}{G'(\omega)^2 + G''(\omega)^2} \quad (9)$$

where $d$ is a numerical constant.

Figure 5 shows the spectrum obtained with a rubber sample. The sample moduli have been independently measured with a rheometer (RDA II) and using the classical time temperature superposition principle.[17] They are shown as a function of frequency in the inset of Fig. 5. Within the frequency range of concern, the elastic modulus $G'$ is nearly constant whereas the loss modulus increases with frequency. According to those values, the critical frequency is $\omega'_1/2\pi \approx 100\text{kHz}$, which is larger than the highest frequency reached in the experiment, the measurements are therefore made in the elastic mode regime.

As shown in Fig. 5, we find a very good agreement between the experimental data and the expected variations of $S(\omega)$ given by equation (9) and using the rheological measurements of the moduli. The agreement moreover extends over nearly five decades in frequency. SFSR is therefore a promising technique for probing viscoelastic media.

**VI CONCLUSION**

In conclusion, we present in this paper a new technique to measure thermally induced fluctuations of a free surface. We measure the intensity of a laser beam that is specularly reflected from that free surface, and we show that the fluctuations of intensities we detect can be related to the bulk properties of either viscous liquids or a viscoelastic medium. Although SFSR is quite simple in its principle and experimental realisation, it provides a very sensitive way to measure fluctuations of media exhibiting very different properties; we moreover show that the computed expression of the signal is in remarkable agreement with the experimental spectra. As other optical techniques, it is non-invasive and non-perturbative of thermodynamic equilibrium; it moreover offers both excellent spatial and temporal resolutions. It should constitute a well adapted technique to study complex fluids, and in particular to probe dynamical phenomena in those fluids.

**ACKNOWLEDGEMENT**

We thank Hélène Montès and Hugues Bodiguel who performed the rheological measurements on the rubber sample.

**APPENDIX: DERIVATION OF $S(\omega)$**

Let $\vec{Q}$ be the transfer wave vector of the reflected beam, $\vec{Q} = \vec{q}_{col} - \vec{q}_{inc}$, where $\vec{q}_{inc}$ and $\vec{q}_{col}$ are respectively the wave vector of the incident beam and the one for the detected intensity. We will remain here in the frame of a nearly normal reflection such that $\vec{q}_{inc}$, is normal to the plane of the surface at rest. Thus the vertical component of $\vec{Q}$ is constant. Let us call $\vec{q}$ the component of $\vec{Q}$ in the plane of the interface. Hence, using the notations of Fig. 1, $\vec{q} = 2\pi \sin\theta/\lambda \, (\cos\varphi, \sin\varphi)$. Finally, let $E(r)$ be the amplitude of the electric field on the surface. We also assume that the beam is Gaussian, and that its beam waist coincides with the interface, hence $E(r)$ is given by:

$$E(r) = E_0 e^{-\frac{r^2}{2R^2}} \quad (A1)$$

where R is the beam size at beam waist.

The intensity $i_c$ of the light collected for a given wave vector $\vec{q}$ is given by the squared modulus of the electric field, i.e. :

$$i_c = \iint d\vec{r} \, d\vec{r}' E(r) E(r') e^{i\left[\vec{q}.(\vec{r}-\vec{r}') + \frac{4\pi}{\lambda}[h(\vec{r},t) - h(\vec{r}',t)]\right]} \quad (A2)$$

In practice, we measure the intensity of light, respectively in the upper and in the lower part of the beam, at infinity. In the approximation where $h(\vec{r},t) - h(\vec{r}',t)$ is small as compared to $\lambda$, which corresponds to our experimental situation, the exponential can be expanded to the first order. The intensity received by the upper quadrant of the photodiode is the integral of $i_c$ over the half space corresponding to $q_x > 0$ (i.e. $0 < \theta < \pi/2$ and $0 < \varphi < \pi$):

$$I_+ = \frac{I_0}{2} + 2i \int_{q_x>0} d\vec{q} \cos\theta \iint d\vec{r} d\vec{r}' \, E(r) E(r') [h(\vec{r};t) - h(\vec{r}';t)] e^{-i\vec{q}.(\vec{r}-\vec{r}')} \quad (A3)$$

$I_0$ is the total intensity of the beam after reflection. The first integral is computed over the upper part of the beam only. The right hand term contains two parts, respectively in $h(\vec{r};t)$ and $-h(\vec{r}';t)$ that are complex conjugated. Using the fact that $E(r)$ has a cylindrical symmetry and introducing the notations:

$$\hat{E}(k) = \frac{1}{A}\int d\vec{r}E(r)e^{i\vec{k}.\vec{r}} = 2\pi R^2 E_0 e^{-k^2R^2/2} \tag{A4}$$

$$h(\vec{k};t) = \frac{1}{A}\iint d\vec{r}h(\vec{r};t)e^{i\vec{k}.\vec{r}} \tag{A5}$$

*A* being the sample surface area.

Equation (A3) finally yields:

$$I_+ = \frac{I_0}{2} + \frac{A}{4\pi^2}\int d\vec{k}F(\vec{k})h(\vec{k};t) \tag{A6}$$

where:

$$F(\vec{k}) = -\frac{4A^2}{i\pi\lambda}\int_0^{\pi}d\varphi \int_{-\pi/2}^{\pi/2}\hat{E}\left(\frac{2\pi}{\lambda}|\sin\theta|\right)\cos\theta\sin\theta d\theta \int d\vec{k}\hat{E}(|\vec{q}+\vec{k}|) \tag{A7}$$

The measured signal $S(\omega) = \frac{1}{I_0^2}\int_{-\infty}^{\infty}dt\langle\delta I(\tau)\delta I^*(t+\tau)\rangle_\tau e^{-i\omega t}$ therefore writes:

$$S(\omega) = \frac{1}{2\pi I_0^2}\int_{-\infty}^{\infty}d\vec{k}|F(\vec{k})|^2 P_k(\omega) \tag{A8}$$

We denote as $\alpha$ the angle such that $d\vec{k} = kdkd\alpha$. $S(\omega)$ can further be expressed as:

$$S(\omega) = \int_0^{\infty}P_k(\omega)kdk\Phi(k) \tag{A9}$$

where the kernel $\Phi(k)$ is given by:

$$\Phi(k) = \frac{1}{2\pi I_0^2}\int_{\alpha=0}^{2\pi}d\alpha|F(\vec{k})|^2 \tag{A10}$$

Integrating $\Phi(k)$ finally yields:

$$\Phi(k) = \frac{2}{\pi\lambda^2}e^{-\frac{k^2R^2}{2}}\int_0^{2\pi}d\alpha\,\text{erf}^2\left(\frac{kR}{2}\sin\alpha\right) \tag{A11}$$

A good approximation for the latter expression is given by:

$$\Phi(k) \cong b\frac{k^2R^2}{\lambda^2}e^{-\frac{k^2R^2}{c}} \tag{A12}$$

where $b$ and $c$ are numerical factors: $b$=0.96 and $c$=1.68.

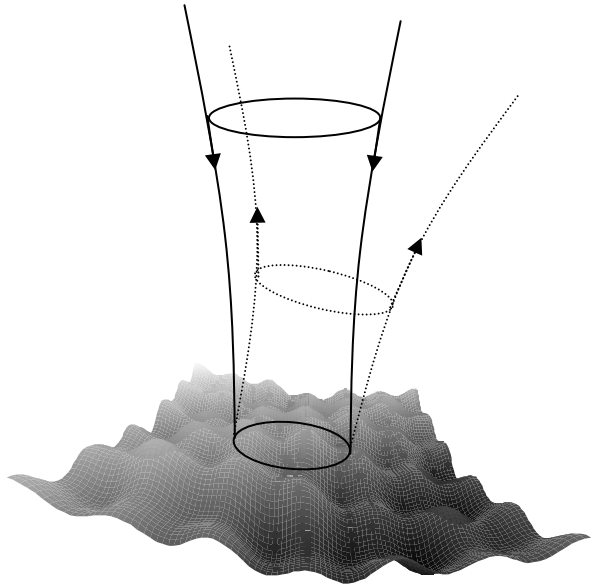

**- a -**

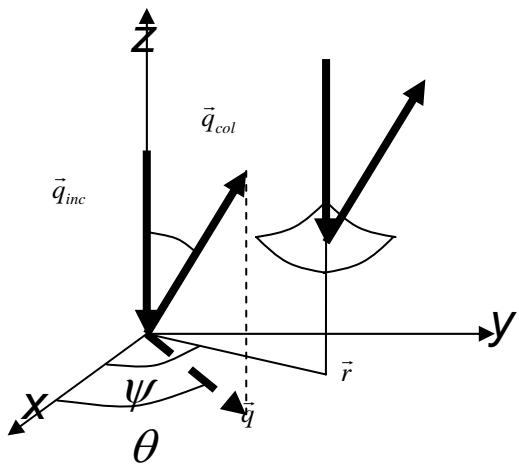

**- b -**

Fig. 1: (a) Principle of the measurement: the position of a laser beam that reflects from a free surface is sensitive to the slope of that surface. Thermally excited waves with wavelength larger than the beam size thus contribute to the deflection of the laser beam. (b) Definition of the adopted notations.

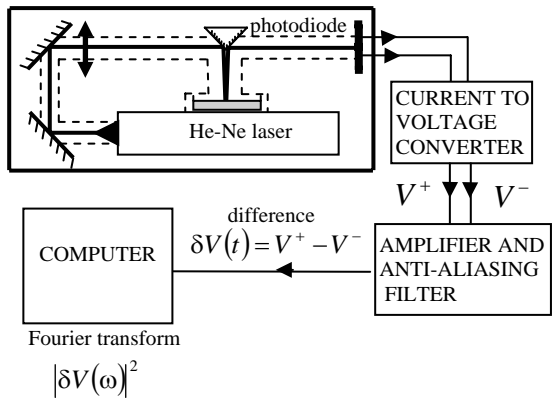

Fig. 2: Schematic representation of the experimental set-up: the laser beam is focused on the sample free surface with a converging lens. The fluctuations in the position of the reflected beam are measured through the fluctuation in the difference of voltage delivered by two quadrants of a photodiode. The signal is amplified, filtered and digitized and its Fourier transform is then computed.

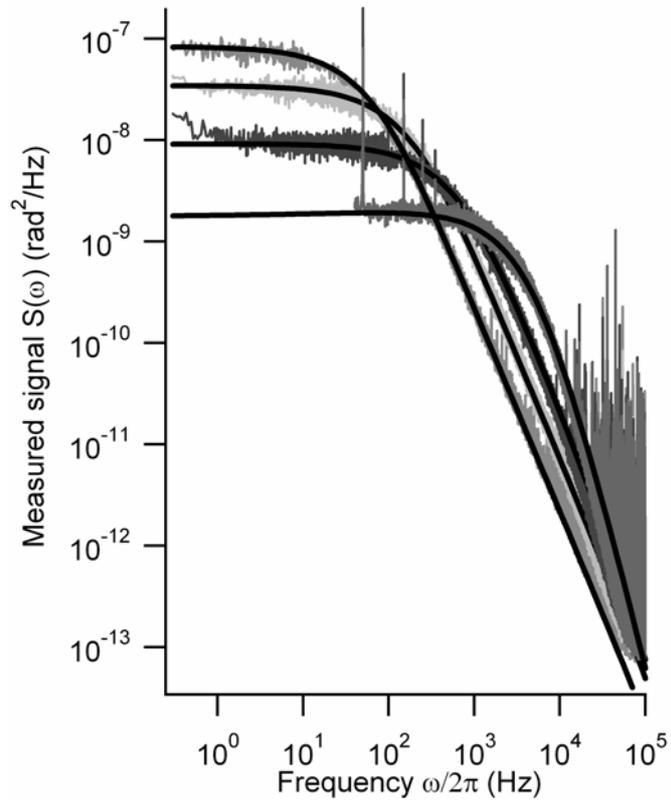

Fig. 3: Spectra measured at the interface air/silicon oil with silicon oils of different viscosities: from top to bottom $\eta$=1 Pa.s, 0.35 Pa.s, 0.1 Pa.s and 0.02 Pas.s. The full black lines represent the theoretical spectra computed from equation (7) for surface tensions of respectively: 21.1mN.m$^{-1}$, 21.1mN.m$^{-1}$, 20.9mN.m$^{-1}$ and 20.1mN.m$^{-1}$.

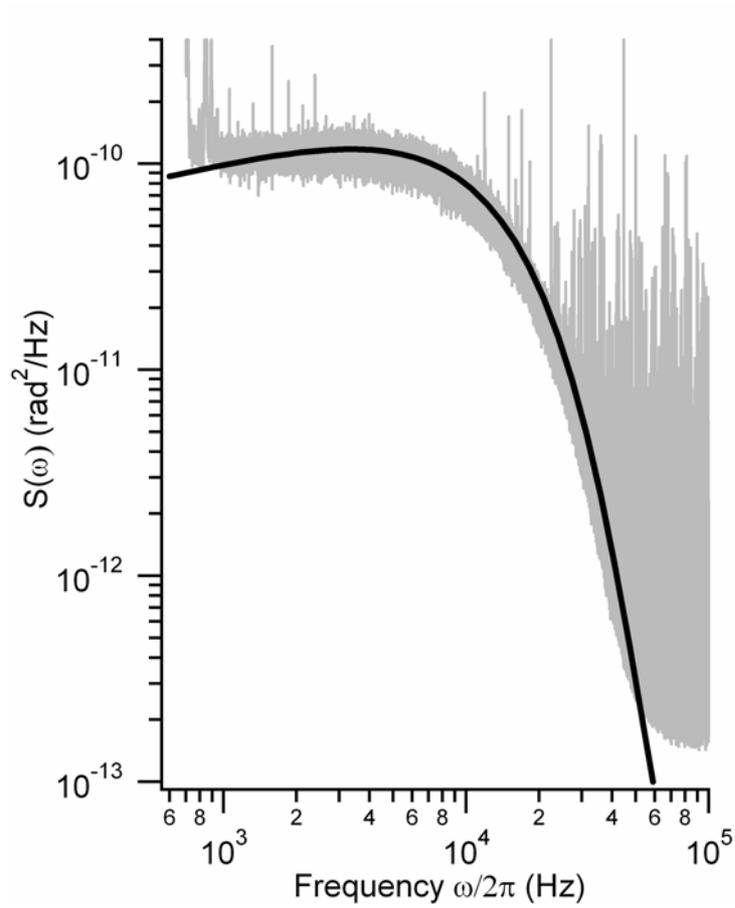

Fig. 4: Spectrum measured at the interface air/water (light grey). The full black line represents the theoretical spectrum computed from equation (7) with viscosity $\eta=1$mPa.s and surface tension $\sigma=60$mN.m$^{-1}$.

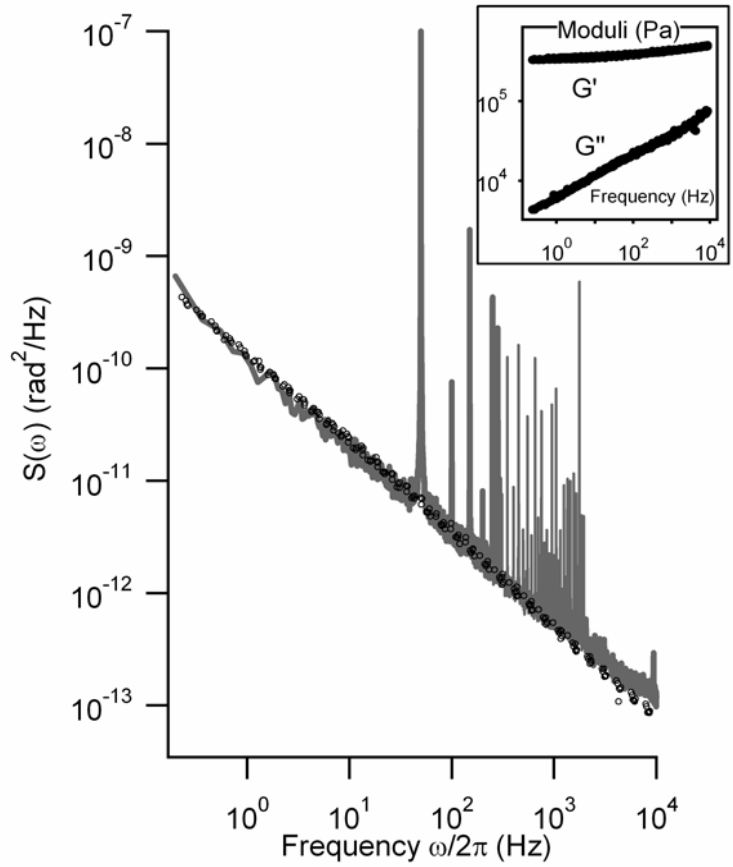

Fig. 5: Spectrum measured at the interface air/rubber: the experimental data (full grey line) is compared with the values of $S(\omega)$ given by equation (9), and using measurements of the moduli (black dots) made with a rheometer and that are shown in the inset. The peaks are caused by parasitic electromagnetic radiations.